\documentclass[runningheads]{llncs}
\usepackage[T1]{fontenc}
\usepackage{graphicx}
\usepackage{array}
\usepackage{amsmath}
\usepackage{stmaryrd}
\usepackage{macros}
\usepackage{makecell}
\usepackage{multirow}
\usepackage{arydshln}
\usepackage{wrapfig}
\usepackage{caption}
\usepackage{subcaption}
\usepackage{amssymb}
\usepackage{hyperref}

\newcommand\blfootnote[1]{%
  \begingroup
  \renewcommand\thefootnote{}\footnote{#1}%
  \addtocounter{footnote}{-1}%
  \endgroup
}

\begin{document}
\title{CorticalFlow$^{++}$: Boosting Cortical Surface Reconstruction Accuracy, Regularity, and Interoperability}
\titlerunning{CorticalFlow$^{++}$}
%
\author{Rodrigo Santa Cruz$^{\dag\ }$\inst{1,2}\orcidID{0000-0002-5273-7296} \and
L\'eo Lebrat$^{\dag\ }$\inst{1,2} \and
Darren Fu\inst{3} \and
Pierrick Bourgeat\inst{1} \and
Jurgen Fripp\inst{1} \and
Clinton Fookes\inst{2}\and
Olivier Salvado\inst{1,2}}

\authorrunning{R. Santa Cruz et al.}
%
\institute{CSIRO \and Queensland University of Technology \and University of Queensland, Australia.}

\maketitle              

\begin{abstract}
The problem of Cortical Surface Reconstruction from magnetic resonance imaging has been traditionally addressed using lengthy pipelines of image processing techniques like FreeSurfer, CAT, or CIVET. These frameworks require very long runtimes deemed unfeasible for real-time applications and unpractical for large-scale studies. Recently, supervised deep learning approaches have been introduced to speed up this task cutting down the reconstruction time from hours to seconds. Using the state-of-the-art CorticalFlow model as a blueprint, this paper proposes three modifications to improve its accuracy and interoperability with existing surface analysis tools, while not sacrificing its fast inference time and low GPU memory consumption. First, we employ a more accurate ODE solver to reduce the diffeomorphic mapping approximation error. Second, we devise a routine to produce smoother template meshes avoiding mesh artifacts caused by sharp edges in CorticalFlow's convex-hull based template. Last, we recast pial surface prediction as the deformation of the predicted white surface leading to a one-to-one mapping between white and pial surface vertices. This mapping is essential to many existing surface analysis tools for cortical morphometry. We name the resulting method CorticalFlow$^{++}$. Using large-scale datasets, we demonstrate the proposed changes provide more geometric accuracy and surface regularity while keeping the reconstruction time and GPU memory requirements almost unchanged.
\end{abstract}
\section{Introduction}
\blfootnote{$\dag$ Equal contribution}
\blfootnote{Our code is made available at: \\ \url{https://bitbucket.csiro.au/projects/CRCPMAX/repos/corticalflow/browse}}
The problem of cortical surface reconstruction (CSR) consists of estimating triangular meshes for the inner and outer cortical surfaces from a magnetic resonance image (MRI). It is a pivotal problem in Neuroimaging and a fundamental task in clinical studies of neurodegenerative diseases~\cite{du2007different} and psychological disorders~\cite{rimol2012cortical}. Traditionally, this problem is tackled by extensive pipelines of handcrafted image processing algorithms~\cite{fischl2012freesurfer,shattuck2002brainsuite,kim2005automatedclasp,macdonald2000CIVET} which are subject to careful hyperparameter tuning (e.g., thresholds, iteration numbers, and convergence criterion) and very long runtimes. 

To overcome these issues, deep learning (DL) based approaches have been proposed recently~\cite{cruz2021deepcsr,henschel2020fastsurfer,ma2021pialnn,lebrat2021corticalflow}. These methods can directly predict cortical surfaces geometrically close to those produced by traditional methods while reducing their processing time from hours to seconds. More specifically, the current state-of-the-art DL model for CSR, named CorticalFlow~\cite{lebrat2021corticalflow}, consists of a sequence of diffeomorphic deformation modules that learn to deform a template mesh towards the target cortical surface from an input MRI. This method takes only a few seconds to produce cortical surfaces with sub-voxel accuracy.

In this paper, we propose three modifications to improve the accuracy of CorticalFlow and its interoperability to other surface processing tools without increasing its reconstruction time and GPU memory consumption.
First, we upgrade the Euler method used by CorticalFlow to solve the flow ODE responsible for computing the diffeomorphic mapping to the fourth-order Runge-Kutta method~\cite{press1992runge}. This tool provides more accurate ODE solutions leading the model to better approximate the target surfaces and reducing the number of self-intersecting faces on the reconstructed meshes.
Second, instead of using the convex hull of the training surfaces as a template, we propose a simple routine to generate smooth templates that tightly wrap the training surfaces. This new template eases the approximation problem leading to more accurate surfaces while suppressing mesh artifacts in highly curved regions. 
Finally, inspired by Ma
et al.~\cite{ma2021pialnn}, we leverage the estimated white surfaces as the initial mesh template for learning and predicting the pial surfaces. This approach provides a better ``starting point'' to the approximation problem as well as a one-to-one mapping between white and pial surface vertices which facilitates the use of the generated surfaces in existing surface-based analysis tools~\cite{schaer2008surface,fischl1999cortical}. In acknowledgment of the CorticalFlow framework, we name our resulting method CorticalFlow$^{++}$.

Using a large dataset of MRI images and pseudo-ground-truth surfaces, we compare CorticalFlow and CorticalFlow$^{++}$ performance in the reconstruction of cortical surfaces from MRI on three perspectives: geometric accuracy, mesh regularity, and time and space complexity for the surface reconstruction. We conclude that the proposed CorticalFlow$^{++}$ improves upon CorticalFlow, on average, by 19.11\% in terms of Chamfer distance and 56.77\% in terms of the percentage of self-intersecting faces. Additionally, it adds only half a second to the final surface reconstruction time while keeping the same GPU memory budget. 

\section{Related Work}

Traditional cortical surface reconstruction frameworks like FreeSurfer\cite{fischl2012freesurfer}, BrainSuite~\cite{shattuck2002brainsuite}, and CIVET~\cite{macdonald2000CIVET} involve two major steps usually accomplished by lengthy sequences of image processing techniques. They first voxel-wise segment the input MRI, then fit surfaces enclosing the gray matter tissue delimiting the brain cortex. More specifically, the widely used FreeSurfer V6 framework for cortical surface analysis from MRI uses an atlas-based segmentation~\cite{fischl2002whole} and a deformable model \cite{dale1999cortical} for surface fitting on these segmented volumes. Recently, Henschel et al.~\cite{henschel2020fastsurfer} accelerated this framework with a modern DL brain segmentation model and a fast spectral mesh processing algorithm for spherical mapping, cutting down FreeSurfer's processing time to one hour.

In contrast, supervised deep learning approaches leverage large datasets of MRIs and pseudo-ground-truth produced with traditional methods to train high-capacity neural networks to predict surfaces directly from the MRI. DeepCSR~\cite{cruz2021deepcsr} trains an implicit surface predictor and CorticalFlow~\cite{lebrat2021corticalflow} learns a diffeomorphic deformable model. Similarly, PialNN~\cite{ma2021pialnn} also learns a deformable model but focuses only on pial surface reconstruction, receiving as input the white matter surface generated with traditional methods. 

Building upon the success and the powerful framework of CorticalFlow, this paper proposes to address its main limitations, aiming to improve its accuracy and interoperability with existent surface analysis tools, but without severely degrading its inference time and GPU memory consumption. 

\section{Method}
In this Section, we present the proposed method CorticalFlow$^{++}$. We start by reviewing the original CorticalFlow framework and its main components, then we introduce our proposed changes. 

\subsection{CorticalFlow Framework \label{sec:cf}}
\begin{figure}[t]
    \centering
     \begin{subfigure}[b]{0.35\textwidth}
         \centering
         \includegraphics[width=\textwidth]{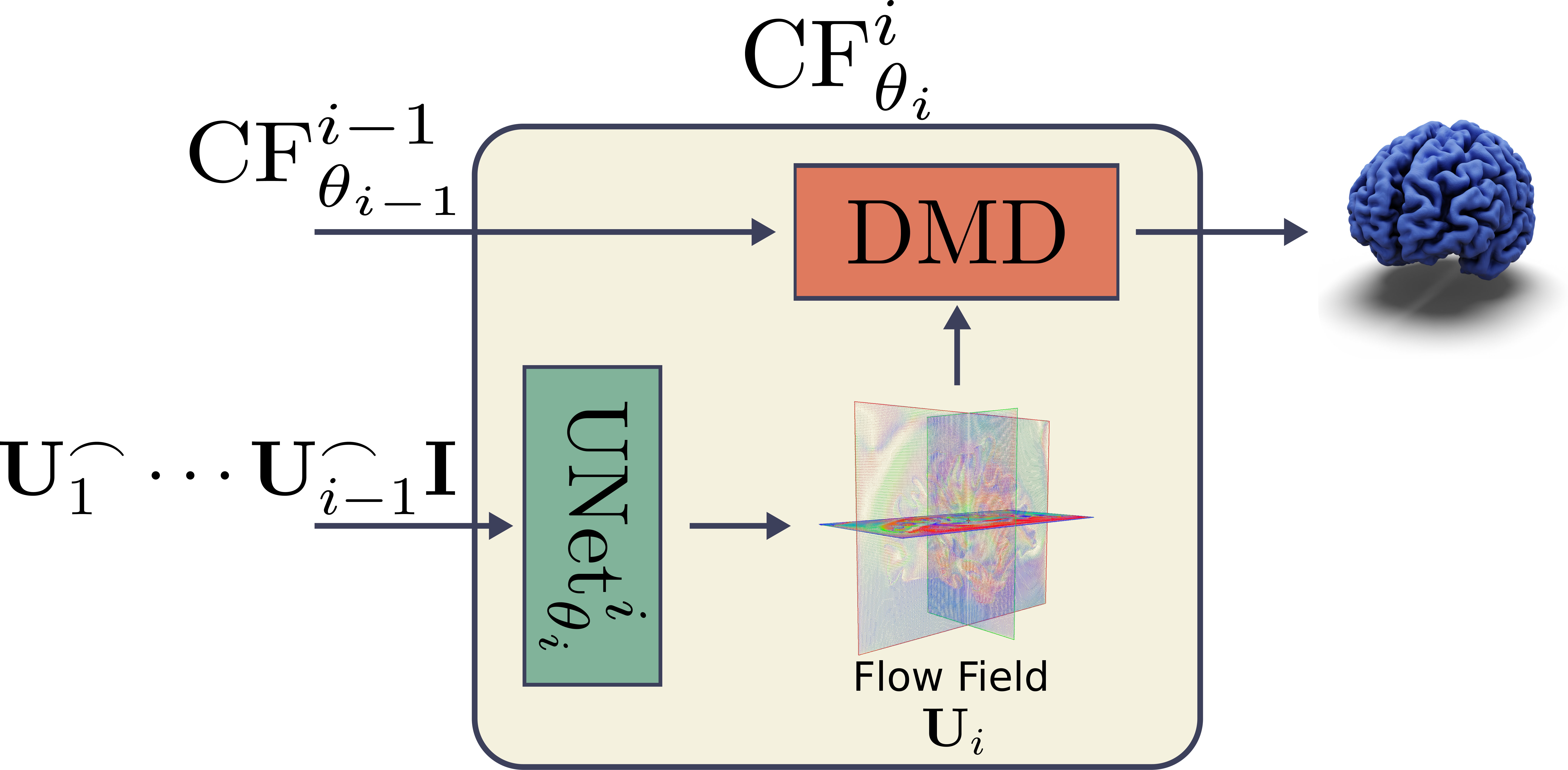}
         \caption{}
         \label{fig:cf_def_block}
     \end{subfigure}
     \hfill
     \begin{subfigure}[b]{0.15\textwidth}
         \centering
         \includegraphics[width=\textwidth]{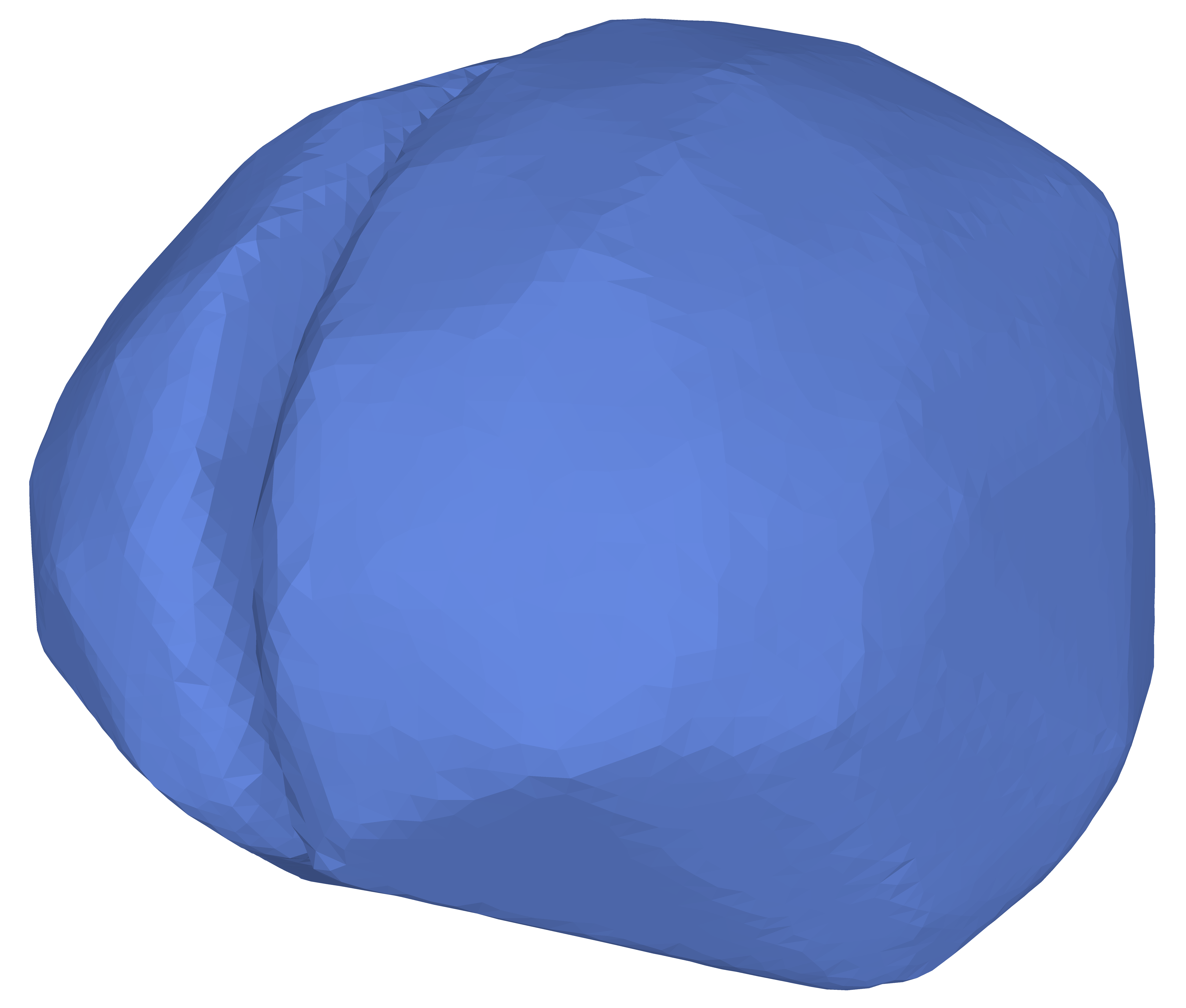}
         \caption{}
         \label{fig:tpl_neurips}
     \end{subfigure}
     \hfill
     \begin{subfigure}[b]{0.15\textwidth}
         \centering
         \includegraphics[width=\textwidth]{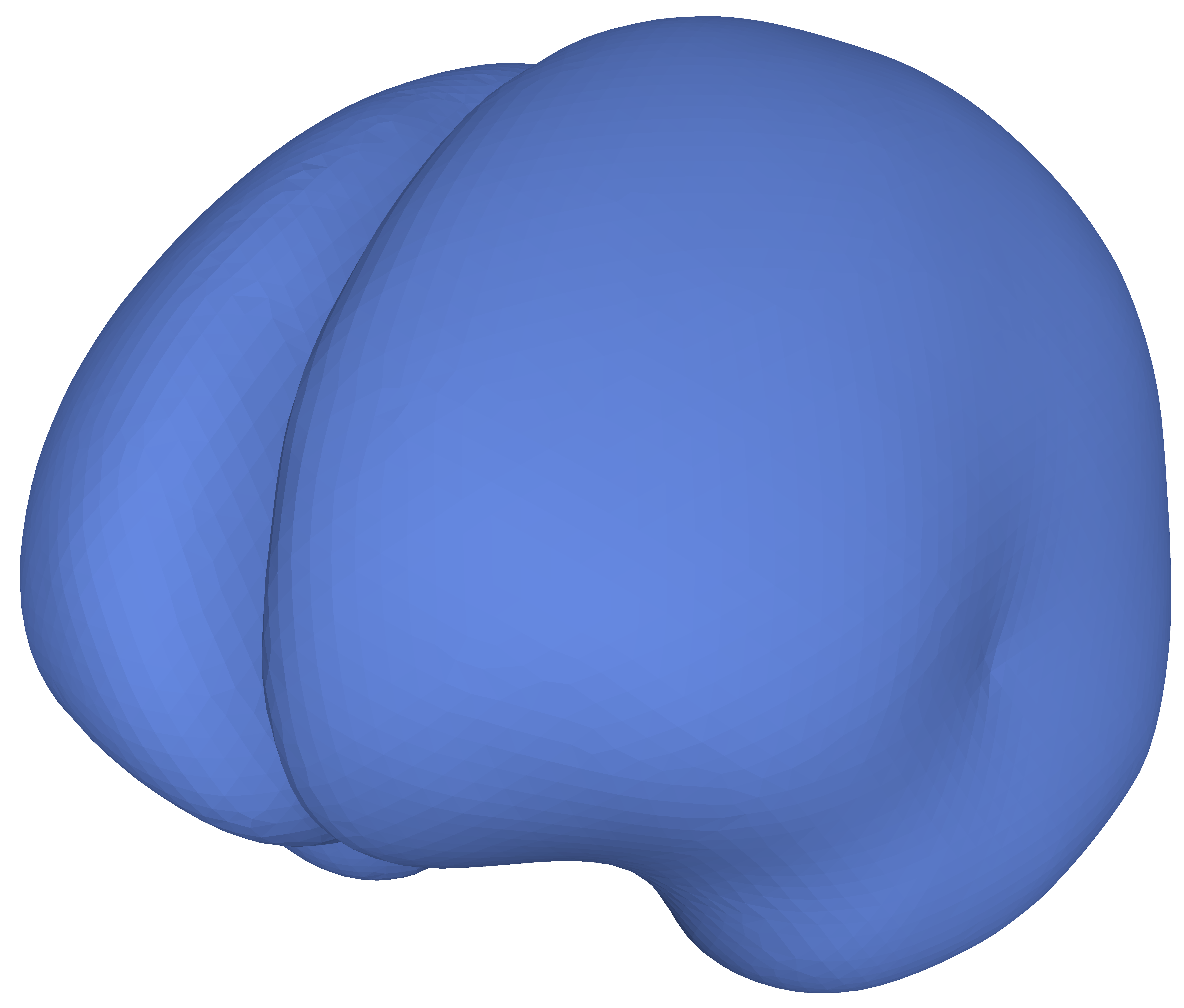}
         \caption{}
         \label{fig:tpl_miccai}
     \end{subfigure}
     \hfill
     \begin{subfigure}[b]{0.25\textwidth}
         \centering
         \includegraphics[width=\textwidth]{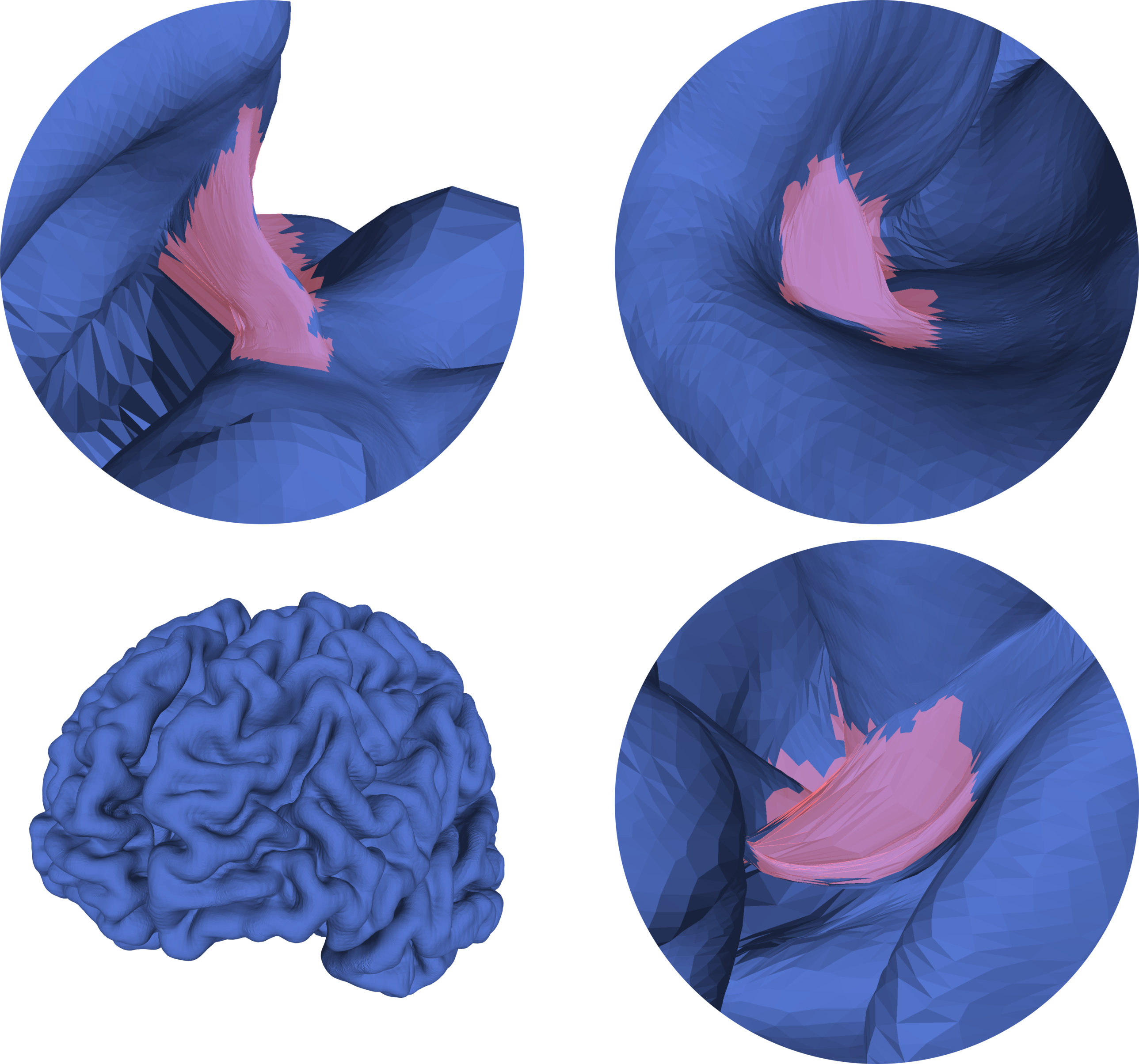}
         \caption{}
         \label{fig:mesh_defects}
     \end{subfigure}
        \caption{(\subref{fig:cf_def_block}) CorticalFlow's Deformation block. (\subref{fig:tpl_neurips}) Convex-hull based CorticalFlow's mesh template. (\subref{fig:tpl_miccai}) CorticalFlow$^{++}$'s proposed template. (\subref{fig:mesh_defects}) Examples of mesh artifacts caused by sharp edges in CorticalFlow's template.\label{fig:method}}
\end{figure}

Originally proposed by Lebrat et al.~\cite{lebrat2021corticalflow}, CorticalFlow consists of a chain of deformation blocks that receives as input a 3-dimensional MRI $\mathbf{I} \in \mathbb{R}^{H\times W \times D}$ and deforms an initial template mesh $\mathcal{T}$ producing the desired cortical surface mesh. 
As shown in \figref{fig:cf_def_block}, a CorticalFlow's deformation block comprises a U-Net~\cite{ronneberger2015unet} and a Diffeomorphic Mesh Deformation module (DMD)~\cite{lebrat2021corticalflow}. 
The U-Net in the $i$-th deformation block, denoted as $\text{UNet}^{i}_{\theta_i}$ and parametrized by $\theta_i$, outputs a stationary flow field $\mathbf{U}_i \in \mathbb{R}^{H\times W \times D \times 3}$, while it receives as input the channel-wise concatenation of the input MRI and all the previous blocks' U-Nets outputs $\lbrace \mathbf{U}_1, \ldots, \mathbf{U}_{i-1} \rbrace$. 
The predicted flow field encodes how the template mesh should be deformed to approximate the target cortical surface.
The DMD module receives as input the block's UNet predicted flow field $\mathbf{U}_i$ and computes a diffeomorphic mapping $\Phi :[0,1] \times \mathbb{R}^3 \rightarrow \mathbb{R}^3$ for each vertex position $\mathbf{x} \in \mathbb{R}^3$ in the resulting mesh computed by the previous deformation block. Formally, the DMD module solves the flow ODE,
\begin{equation}\label{eqn:autoODE}
\frac{\text{d}\Phi(s;\mathbf{x})}{\text{d}s} = \mathbf{U}\left(\Phi(s;\mathbf{x})\right), \text{ with } \Phi(0;\mathbf{x}) = \mathbf{x},
\end{equation}
using the forward Euler method~\cite{euler1794institutiones}. 
This flow ODE formulation preserves the topology of the initial template mesh without producing self-intersecting faces. 
Hence, CorticalFlow with $k$ deformations ($\text{CF}^k$) can be written using the following recurrence,
\begin{align}\label{eqn:RecCF}
    \text{CF}^{1}_{\theta_1}(\mathbf{I},\mathcal{T}) &= \text{DMD}(\text{UNet}^{1}_{\theta_1}(\mathbf{I}),\mathcal{T})) \nonumber \\
    \text{CF}^{i+1}_{\theta_{i+1}}(\mathbf{I},\mathcal{T}) &= \text{DMD}(\text{UNet}^{i+1}_{\theta_{i+1}}(\mathbf{U}_1^\frown \cdots \mathbf{U}_{i}^\frown\mathbf{I}),\text{CF}^{i}_{\theta_{i}}(\mathbf{I},\mathcal{T})) \quad \text{for 
    } i \geq 1, 
\end{align}
with $\mathbf{A}^{\frown}\mathbf{B}$ denoting the channel-wise concatenation of the tensors $\mathbf{A}$ and $\mathbf{B}$. As in \cite{lebrat2021corticalflow}, we focus on CorticalFlow with three deformation blocks ($\text{CF}^3$). 

In order to train $\text{CF}^3$, the authors propose a multi-stage approach where the deformation blocks are trained successively keeping previous blocks' weights frozen and using template meshes of increasing resolution $\mathcal{T}_{i}$. Mathematically, at each stage $i$, they optimize the following objective,
\begin{equation}\label{eqn:problem}
\argmin_{\theta_i} \sum_{(\textbf{I},\textbf{S})\in \mathcal{D}} \mathcal{L} \big(\text{CF}^{i}_{\theta_{i}}(\mathbf{I},\mathcal{T}_{i}),S),
\end{equation}
where $\mathcal{D}$ is a dataset composed of pairs of MRIs and their respective triangle meshes $S$ representing some cortical surface. $\mathcal{L}(\cdot, \cdot)$ is the training loss composed of mesh edge loss~\cite{wang2018pixel2mesh} and Chamfer Distance loss \cite{wang2018pixel2mesh}. Note that for each cortical surface and brain hemisphere, a separate CorticalFlow is trained independently. 

\subsection{Higher Order ODE Solver}
DMD modules in CorticalFlow solve the flow ODE (\ref{eqn:autoODE}) using the forward Euler method which consists of an iterative method defined by the following integration step:
\begin{equation}\label{eqn:forwardEuler}
\hat{\Phi}(h,\mathbf{x})= \mathbf{x} + h\mathbf{U}(\mathbf{x}),
\end{equation}
where $h$ is the algorithm step-size and $\mathbf{U}(\mathbf{x})$ is the linear interpolation of a predicted flow field $\mathbf{U}$ at mesh vertex position $\mathbf{x}$. $\hat{\Phi}$ is a numerical approximation of the true diffeomorphic mapping $\Phi$ due to the interpolation and discretization errors inherent to this application. 

Since the Euler method only provides an approximation of the continuous problem with an error that decreases linearly as $h$ decreases, it may require a large number of integration steps to approximate accurately the continuous solution. Otherwise, the resulting mapping may cease to be invertible. For these reasons, we propose to use the Runge-Kutta~\cite{press1992runge} explicit fourth-order approximation method also known as $RK4$. The integration step of this method consists of the weighted average of four slopes estimated at the beginning, two different midpoints, and end of the step size interval. For our stationary vector field, the $RK4$ integration step is defined as,
\begin{equation}\label{eqn:rk4}
\hat{\Phi}(h,\mathbf{x}) = \mathbf{x} + \frac{1}{6} \left[ k_1 + 2k_2 + 2k_3 + k_4\right],
\end{equation}
where $k_1 = \mathbf{U}\left(\mathbf{x}\right)$, $k_2 = \mathbf{U}\left(\mathbf{x} + h\frac{k_1}{2}\right)$, $k_3 = \mathbf{U}\left(\mathbf{x} + h\frac{k_2}{2}\right)$, and $k_4= \mathbf{U}\left(\mathbf{x} + hk_3\right)$ are the averaged slopes.

\subsection{Smooth Templates \label{sec:new_tpl}}
CorticalFlow's template mesh consists of the convex-hull of all cortical surface meshes in the training set. This approach has two main shortcomings:
\begin{enumerate}
    \item Even target surfaces with small differences between them can lead to a ``loose'' template. Consequently, the model has to learn ``large'' deformations making the smooth approximation problem harder. For pial surfaces, this problem is even worse because some template regions may lay outside the image bounds where the predicted flow field is undefined.
    \item The convex-hull is defined by a set of intersecting planes which leads to sharp edges as shown in \figref{fig:tpl_neurips}. These edges are very hard to be unfolded by a smooth deformable model like CorticalFlow, because it requires non-smooth deformations with drastic local changes of direction in its flow field representation. Hence, these edges may remain in the predicted mesh as undesirable artifacts (see \figref{fig:mesh_defects}).
\end{enumerate}

To overcome these issues, we develop genus zero smooth mesh templates that tightly wrap all training meshes. We first compute a signed distance map for every target surface mesh in the training set by computing the largest 3D bounding box that contains these meshes, create $512^3$ voxel-grids into this bounding box, and populate these voxel-grids with the signed distance to each target mesh. These signed distance maps are implicit representations of the target meshes where voxels with positives values are outside of the mesh and voxels with negative values are inside of the mesh. Then, by thresholding the binary union of these maps and running the standard marching cubes algorithm~\cite{lorensen1987marching}, we obtain a template mesh that is very tight around all training surfaces. However, this template mesh looks ``blocky'' with many small sharp edges and undesired topological defects. The template mesh is thus smoothed using the Laplacian smoothing algorithm~\cite{herrmann1976laplacian,sorkine2004laplacian} and re-meshed with Delaunay triangulation~\cite{bobenko2007discrete}. The result is a smooth template mesh with spherical topology tightly wrapping all training set surfaces (see \figref{fig:tpl_miccai}). Finally, we apply a topology preserving mesh subdivision algorithm~\cite{meshlab} to generate template meshes at different resolutions which are required to train CorticalFlow. Implementations of the used algorithms are available in the Libigl~\cite{libigl} and MeshLab~\cite{meshlab} toolboxes.

\subsection{White To Pial Surface Morphing}
In Lebrat et al.~\cite{lebrat2021corticalflow}, a separate CorticalFlow model is trained for each cortical surface (i.e., pial and white) and each brain hemisphere (i.e., left and right). This approach leads to reconstructed surface meshes without a one-to-one mapping between the vertices in the white and pial surfaces on the same brain hemisphere. In the absence of this mapping, many existent surface analysis tools can not process the generated surfaces. Additionally, as also observed in Ma et al.~\cite{ma2021pialnn}, the pial surface may only differ from the white surface by a ``small'' deformation thanks to the natural anatomical agreement between these surfaces. Therefore, we propose to predict pial surfaces by learning to deform the predicted white surfaces instead of using a pial surface template mesh. 

Formally, the resulting model for pial surfaces with $k$ deformation blocks ($\text{CFP}^k$) can be restated by the following recurrence,
\begin{align}\label{eqn:RecCFPial}
    \text{CFP}^{1}_{\theta_1}(\mathbf{I}, \mathcal{T}^{w}) &= \text{DMD}(\text{UNet}^{1}_{\theta_1}(\mathbf{I}),\text{CFW}^{k'}(\mathbf{I}, \mathcal{T}^{w})) \nonumber \\
    \text{CFP}^{i+1}_{\theta_{i+1}}(\mathbf{I},\mathcal{T}^{w}) &= \text{DMD}(\text{UNet}^{i+1}_{\theta_{i+1}}(\mathbf{U}_1^\frown \cdots \mathbf{U}_{i}^\frown\mathbf{I}),\text{CFP}_{i}(\mathbf{I},\mathcal{T}^{w})) \quad \text{for 
    } i \geq 1, 
\end{align}
where $\text{CFW}^{k'}$ is a CorticalFlow model with $k'$ deformation blocks pretrained to reconstruct the same hemisphere white surface as described in \secref{sec:cf} and $\mathcal{T}^{w}$ is its respective template mesh for white surfaces generated as described in \secref{sec:new_tpl}. Note that the formulation of $\text{CFW}^{k'}$ remains the one stated in \eqnref{eqn:RecCF} and we only use template meshes for the white surfaces since the pial surfaces are obtained by deforming the predicted white surface. 

\section{Experiments}
\begin{table}[t!]
\centering
\resizebox{\textwidth}{!}{\begin{tabular}{l|cccc|cccc|}
\hline
& \multicolumn{4}{c|}{\large{Left Pial Surface}} & \multicolumn{4}{c|}{\large{Right Pial Surface }} \\
& CH($mm$) $\downarrow$ & HD($mm$) $\downarrow$ & CHN $\uparrow$ & \% SIF $\downarrow$ & CH($mm$) $\downarrow$ & HD($mm$) $\downarrow$ & CHN $\uparrow$ & \% SIF $\downarrow$ \\
 \hline
 \makecell{\large{CorticalFlow} \\  3.22 sec / 2.82 GB } & \makecell{$0.681$ \\ $(\pm 0.098)$} & \makecell{$0.802$ \\ $(\pm 0.049)$} & \makecell{$0.932$ \\ $(\pm 0.006)$} & \makecell{$0.686$ \\ $(\pm 0.469)$} & \makecell{$0.693$ \\ $(\pm 0.091)$} & \makecell{$0.815$ \\ $(\pm 0.046)$} & \makecell{$0.929$ \\ $(\pm 0.006)$} & \makecell{$1.239$ \\ $(\pm 0.629)$} \\
 
 \hdashline[5pt/5pt]
 
 \makecell{\large{CorticalFlow + RK4} \\  3.55 sec / 2.82 GB } &  \makecell{$0.629$ \\ $(\pm 0.100)$} & \makecell{$0.761$ \\ $(\pm 0.042)$} & \makecell{$0.937$ \\ $(\pm 0.006)$} & \makecell{$0.502$ \\ $(\pm 0.196)$} & \makecell{$0.580$ \\ $(\pm 0.082)$} & \makecell{$0.751$ \\ $(\pm 0.038)$} & \makecell{$0.943$ \\ $(\pm 0.006)$} & \makecell{$0.280$ \\ $(\pm 0.133)$}  \\
 
 
 \makecell{\large{CorticalFlow + W2P} \\ 3.63 sec  / 2.82 GB } & \makecell{$0.545$ \\ $(\pm 0.082)$} & \makecell{$0.730$ \\ $(\pm 0.037)$} & \makecell{$0.943$ \\ $(\pm 0.006)$} & \makecell{$0.188$ \\ $(\pm 0.116)$} & \makecell{$0.540$ \\ $(\pm 0.075)$} & \makecell{$0.729$ \\ $(\pm 0.033)$} & \makecell{$0.945$ \\ $(\pm 0.006)$} & \makecell{$0.176$ \\ $(\pm 0.134)$} \\
 
 \hdashline[5pt/5pt]
 
\makecell{\large{PialNN$^\dagger$} \\  white gen. + 0.880 secs / 1.92 GB } & \makecell{$5.500$ \\ $(\pm 0.786)$} & \makecell{$2.793$ \\ $(\pm 0.220)$} & \makecell{$0.792$ \\ $(\pm 0.009)$} & \makecell{$4.730$ \\ $(\pm 0.841)$} & \makecell{$5.948$ \\ $(\pm 0.811)$} & \makecell{$2.898$ \\ $(\pm 0.212)$} & \makecell{$0.789$ \\ $(\pm 0.011)$} & \makecell{$4.537$ \\ $(\pm 0.815)$}  \\

\makecell{\large{PialNN$^\star$} \\  white gen. + 0.880 secs / 1.92 GB } & \makecell{$1.388$ \\ $(\pm 0.223)$} & \makecell{$1.251$ \\ $(\pm 0.120)$} & \makecell{$0.864$ \\ $(\pm 0.011)$} & \makecell{$10.507$ \\ $(\pm 1.908)$} & \makecell{$1.374$ \\ $(\pm 0.217)$} & \makecell{$1.236$ \\ $(\pm 0.115)$} & \makecell{$0.863$ \\ $(\pm 0.011)$} & \makecell{$11.159$ \\ $(\pm 1.930)$}  \\
 
 \hdashline[5pt/5pt]
 
 \makecell{\large{CorticalFlow$^{++}$} \\  3.76 sec / 2.82 GB } & \makecell{$0.529$ \\ $(\pm 0.088)$} & \makecell{$0.721$ \\ $(\pm 0.036)$} & \makecell{$0.946$ \\ $(\pm 0.006)$} &  \makecell{$0.069$ \\ $(\pm 0.060)$} & \makecell{$0.528$ \\ $(\pm 0.074)$} & \makecell{$0.723$ \\ $(\pm 0.031)$} & \makecell{$0.946$ \\ $(\pm 0.005)$} & \makecell{$0.099$ \\ $(\pm 0.093)$} \\
 
\hline
 
& \multicolumn{4}{c|}{\large{Left White Surface}} & \multicolumn{4}{c|}{\large{Right White Surface }} \\
& CH($mm$) $\downarrow$ & HD($mm$) $\downarrow$ & CHN $\uparrow$ & \% SIF $\downarrow$ & CH($mm$) $\downarrow$ & HD($mm$) $\downarrow$ & CHN $\uparrow$ & \% SIF $\downarrow$\\
\hline
 \makecell{\large{CorticalFlow} \\  3.22 sec  / 2.82 GB } & \makecell{$0.608$ \\ $(\pm 0.098)$} & \makecell{$0.785$ \\ $(\pm 0.060)$} & \makecell{$0.941$ \\ $(\pm 0.007)$} & \makecell{$0.033$ \\ $(\pm 0.030)$} & \makecell{$0.599$ \\ $(\pm 0.093)$} & \makecell{$0.783$ \\ $(\pm 0.059)$} & \makecell{$0.942$ \\ $(\pm 0.007)$} & \makecell{$0.030$ \\ $(\pm 0.029)$} \\

\hdashline[5pt/5pt]

\makecell{\large{CorticalFlow + RK4} \\  3.55 sec / 2.82 GB } & \makecell{$0.540$ \\ $(\pm 0.107)$} & \makecell{$0.733$ \\ $(\pm 0.055)$} & \makecell{$0.948$ \\ $(\pm 0.006)$} & \makecell{$0.042$ \\ $(\pm 0.039)$} & \makecell{$0.517$ \\ $(\pm 0.089)$} & \makecell{$0.716$ \\ $(\pm 0.044)$} & \makecell{$0.951$ \\ $(\pm 0.006)$} & \makecell{$0.010$ \\ $(\pm 0.023)$}  \\

\makecell{\large{CorticalFlow + NEWTPL} \\  3.04 sec  / 2.82 GB } & \makecell{$0.598$ \\ $(\pm 0.101)$} & \makecell{$0.780$ \\ $(\pm 0.062)$} & \makecell{$0.942$ \\ $(\pm 0.007)$} & \makecell{$0.030$ \\ $(\pm 0.027)$} & \makecell{$0.558$ \\ $(\pm 0.091)$} & \makecell{$0.747$ \\ $(\pm 0.053)$} & \makecell{$0.945$ \\ $(\pm 0.006)$} & \makecell{$0.104$ \\ $(\pm 0.079)$}  \\

\hdashline[5pt/5pt]

\makecell{\large{CorticalFlow$^{++}$} \\  3.35 sec / 2.82 GB } & \makecell{$0.514$ \\ $(\pm 0.090)$} & \makecell{$0.712$ \\ $(\pm 0.044)$} & \makecell{$0.952$ \\ $(\pm 0.006)$} & \makecell{$0.017$ \\ $(\pm 0.023)$} & \makecell{$0.510$ \\ $(\pm 0.083)$} & \makecell{$0.711$ \\ $(\pm 0.040)$} & \makecell{$0.952$ \\ $(\pm 0.006)$} & \makecell{$0.031$ \\ $(\pm 0.040)$}  \\

\hline
\end{tabular}}
\caption{Cortical Surface Reconstruction Benchmark \cite{cruz2021deepcsr}. $\downarrow$ indicates smaller metric value is better, while $\uparrow$ indicates greater metric value is better.}
\label{tab:benchmark}
\end{table}

We now evaluate CorticalFlow$^{++}$ in the cortical surface reconstruction problem. 
First, using the CSR benchmark introduced by Santa Cruz et al.~\cite{cruz2021deepcsr}, we quantify the performance impact of each proposed modification separately. 
This benchmark consists of 3,876 MRI images extracted from the Alzheimer’s Disease Neuroimaging Initiative\footnote{Data used in preparation of this article were obtained from the Alzheimer’s Disease Neuroimaging Initiative
(ADNI) database (adni.loni.usc.edu). As such, the investigators within the ADNI contributed to the design
and implementation of ADNI and/or provided data but did not participate in analysis or writing of this report.
A complete listing of ADNI investigators can be found at: \url{http://adni.loni.usc.edu/wp-content/uploads/how_to_apply/ADNI_Acknowledgement_List.pdf}}
 (ADNI)\cite{Jack2008:ADNI} and their respective pseudo-ground-truth surfaces generated with the FreeSurfer V6.0 cross-sectional pipeline. We strictly follow the benchmark's data splits and evaluation protocol. We refer the reader to \cite{cruz2021deepcsr,lebrat2021corticalflow} for full details on this dataset.
As evaluation metrics for measuring geometric accuracy, we use Chamfer distance (CH), Hausdorff distance (HD), and Chamfer normals (CHN). These distances are computed for point clouds of 200k points uniformly sampled from the predicted and target surfaces. As a measure of surface regularity, we compute the percentage of self-intersecting faces (\%SIF) using PyMeshLab \cite{pymeshlab}. Finally, we also report the average time (in seconds) and the maximum GPU memory (in GB) required by the evaluated methods to reconstruct a cortical surface. \tabref{tab:benchmark} presents these results.

\begin{figure}[t]
    \centering
    \includegraphics[width=\textwidth]{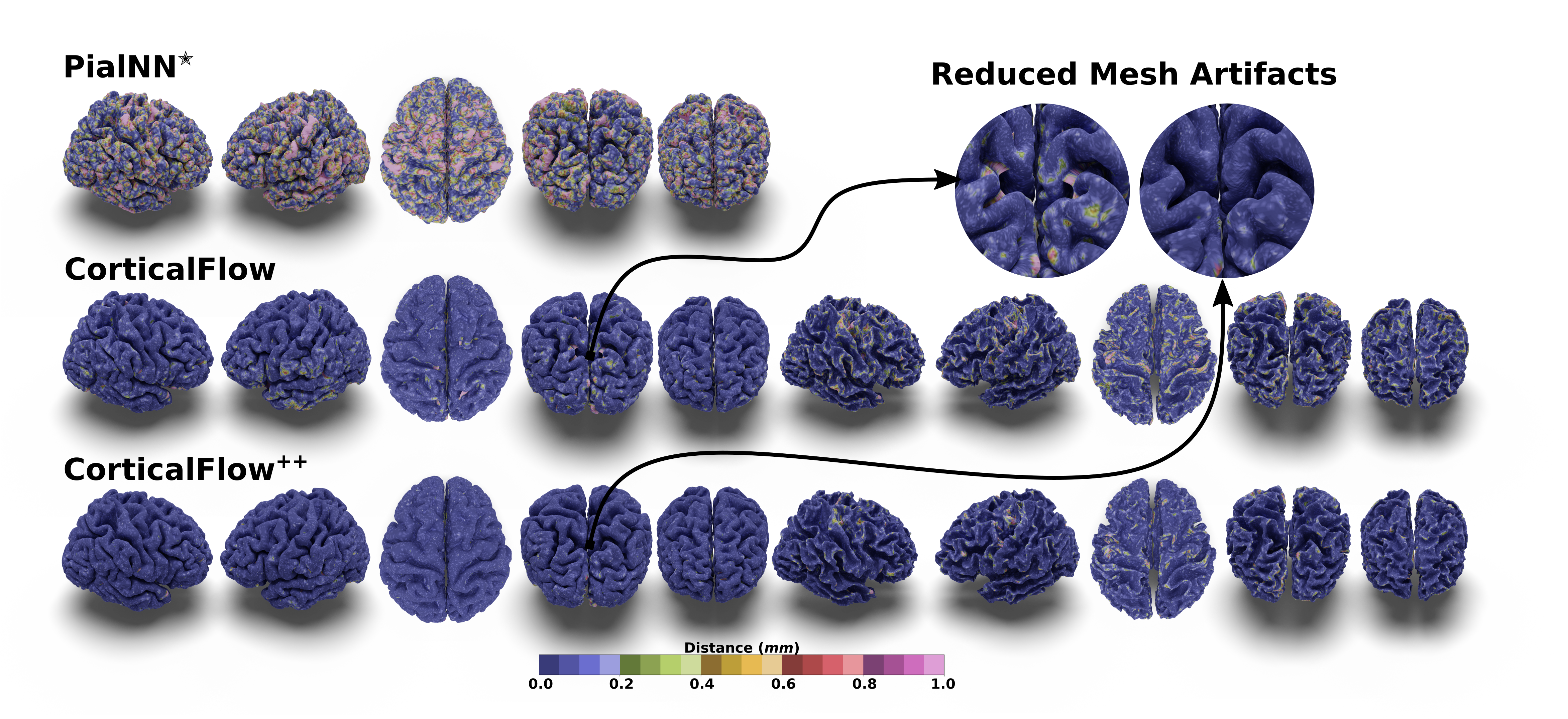}
    \caption{Predicted cortical surfaces color-coded with the distance to the pseudo-ground-truth surfaces. See our supplementary materials for more examples. \label{fig:gen_surfaces}}
\end{figure}

We observe that the adoption of the $RK4$ ODE solver (CorticalFlow + RK4) and the white to pial surface morphing formulation (CorticalFlow + W2P) significantly improves the geometric accuracy and surface regularity of the CorticalFlow baseline. For instance, we noticed an average decrease of 12.2\% and 21.02\% in chamfer distance, respectively. Likewise, the percentage of self-intersecting faces reduces on average by 35.90\% and 79.19\%, respectively. On the other hand, the proposed new templates (CorticalFlow + NEWTPL) present a modest improvement in those criteria, but greatly succeeds in suppressing mesh artifacts as qualitatively shown in \figref{fig:gen_surfaces}. Additionally, none of these changes incur a significant increase in reconstruction time or memory consumption due to their GPU friendly nature. All together, these modifications in CorticalFlow$^{++}$ establish a new state-of-the-art method for cortical surface reconstruction.

\begin{table}[t!]
\centering
\resizebox{\textwidth}{!}{\begin{tabular}{l|cccc|cccc|}
\hline
& \multicolumn{4}{c|}{\large{Left Pial Surface}} & \multicolumn{4}{c|}{\large{Right Pial Surface }} \\
& CH($mm$) $\downarrow$ & HD($mm$) $\downarrow$ & CHN $\uparrow$ & \% SIF $\downarrow$ & CH($mm$) $\downarrow$ & HD($mm$) $\downarrow$ & CHN $\uparrow$ & \% SIF $\downarrow$ \\
 \hline
 \makecell{\large{CorticalFlow} \\  3.22 sec / 2.82 GB } & \makecell{$0.677$ \\ $(\pm 0.099)$} & \makecell{$0.803$ \\ $(\pm 0.056)$} & \makecell{$0.923$ \\ $(\pm 0.007)$} & \makecell{$0.594$ \\ $(\pm 0.319)$} & \makecell{$0.724$ \\ $(\pm 0.106)$} & \makecell{$0.845$ \\ $(\pm 0.063)$} & \makecell{$0.918$ \\ $(\pm0.007)$} & \makecell{$1.467$ \\ $(\pm 0.519)$} \\
 
 \makecell{\large{PialNN$^\dagger$} \\ white gen. + 0.880 secs / 1.92 GB } & \makecell{$5.426$ \\ $(\pm 0.486)$} & \makecell{$2.763$ \\ $(\pm 0.144)$} & \makecell{$0.781$ \\ $(\pm 0.009)$} & \makecell{$4.281$ \\ $(\pm 0.709)$} & \makecell{$5.944$ \\ $(\pm 0.503)$} & \makecell{$2.873$ \\ $(\pm 0.139)$} & \makecell{$0.777$ \\ $(\pm 0.009)$} & \makecell{$4.033$ \\ $(\pm 0.671)$} \\
 
 \makecell{\large{PialNN$^\star$} \\ white gen. + 0.880 secs / 1.92 GB } & \makecell{$1.307$ \\ $(\pm 0.202)$} & \makecell{$1.243$ \\ $(\pm 0.119)$} & \makecell{$0.860$ \\ $(\pm 0.013)$} & \makecell{$9.661$ \\ $(\pm 1.604)$} & \makecell{$1.264$ \\ $(\pm 0.194)$} & \makecell{$1.211$ \\ $(\pm 0.117)$} & \makecell{$0.857$ \\ $(\pm 0.013)$} & \makecell{$10.482$ \\ $(\pm 1.678)$} \\
 
 \makecell{\large{CorticalFlow$^{++}$} \\ 3.76 sec / 2.82 GB } & \makecell{$0.520$ \\ $(\pm 0.082)$} & \makecell{$0.711 $ \\ $(\pm 0.044)$} & \makecell{$0.935$ \\ $(\pm 0.006)$} & \makecell{$0.136$ \\ $(\pm 0.096)$} & \makecell{$0.528$ \\ $(\pm 0.079)$} & \makecell{$0.727$ \\ $(\pm 0.047)$} & \makecell{$0.935$ \\ $(\pm 0.006)$} & \makecell{$0.245$ \\ $(\pm 0.167)$} \\

\hline
\end{tabular}}
\caption{Out-of-train-distribution evaluation on OASIS3 \cite{lamontagne2019oasis}. $\downarrow$ indicates smaller metric value is better, while $\uparrow$ indicates greater metric value is better.}
\label{tab:oasis}
\end{table}

Since the white to pial morphing approach has been previously introduced in \cite{ma2021pialnn}, we compare CorticalFlow$^{++}$ and PialNN~\cite{ma2021pialnn} on the pial surface reconstruction. For this comparison, we report the performance of the publicly available pretrained PialNN model\footnote{https://github.com/m-qiang/PialNN} (PialNN$^\dagger$) as well as training it by ourselves in the CSR benchmark training dataset (PialNN$^\star$). CorticalFlow$^{++}$ compares favourably to both PialNN variants and importantly, it does not need traditional methods to generate white surfaces.

Finally, we extend our evaluation to an out-of-training-distribution dataset. More specifically, we use the trained models from the previous experiment to reconstruct cortical surfaces for a subset of MRIs extracted from the OASIS3 dataset \cite{lamontagne2019oasis}. These generated surfaces are also compared to FreeSurfer V6.0 pseudo-ground-truth surfaces using the same evaluation metrics described above. As shown in \tabref{tab:oasis}, CorticalFlow$^{++}$ significantly outperforms CorticalFlow and PialNN, while presenting comparable surface reconstruction runtime and GPU memory consumption.

\section{Conclusion}
This paper tackles some limitations of CorticalFlow, the current state-of-the-art model for Cortical surface reconstruction from MRI, in order to improve its accuracy, regularity, and interoperability without sacrificing its computational requirements for inference (reconstruction time and maximum GPU memory consumption).
The resulting method,  CorticalFlow$^{++}$,  achieves state-of-the-art performance on geometric accuracy and surface regularity while keeping the GPU memory consumption constant and adding less than a second to the entire surface reconstruction process.

\section{Compliance with Ethical Standards}
This research was approved by CSIRO ethics 2020 068 LR.

\section{Acknowledgements}
This work was funded in part through an Australian Department of Industry, Energy and Resources CRC-P project between CSIRO, Maxwell Plus and I-Med Radiology Network.

\newpage
\section*{Supplementary Material}

\begin{figure}[h!]
    \centering
    \includegraphics[width=0.92\textwidth]{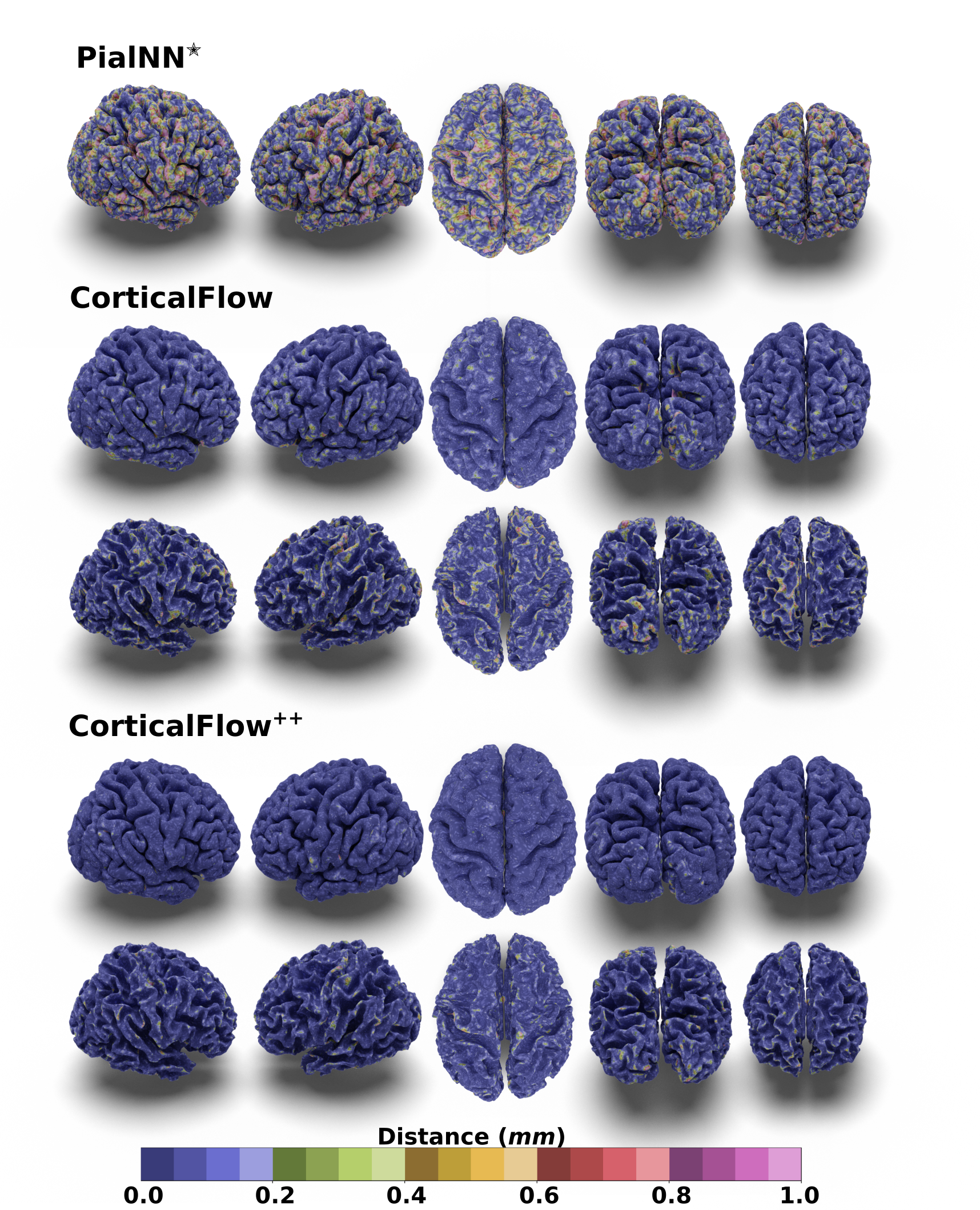} 
    \caption{Predicted cortical surfaces for subject \textbf{099\_S\_0551\_m36} in \textbf{ADNI} dataset. The surfaces are color-coded with the distance to the pseudo-ground-truth surfaces.}
\end{figure}

\begin{figure}[h!]
    \centering
    \includegraphics[height=\textheight]{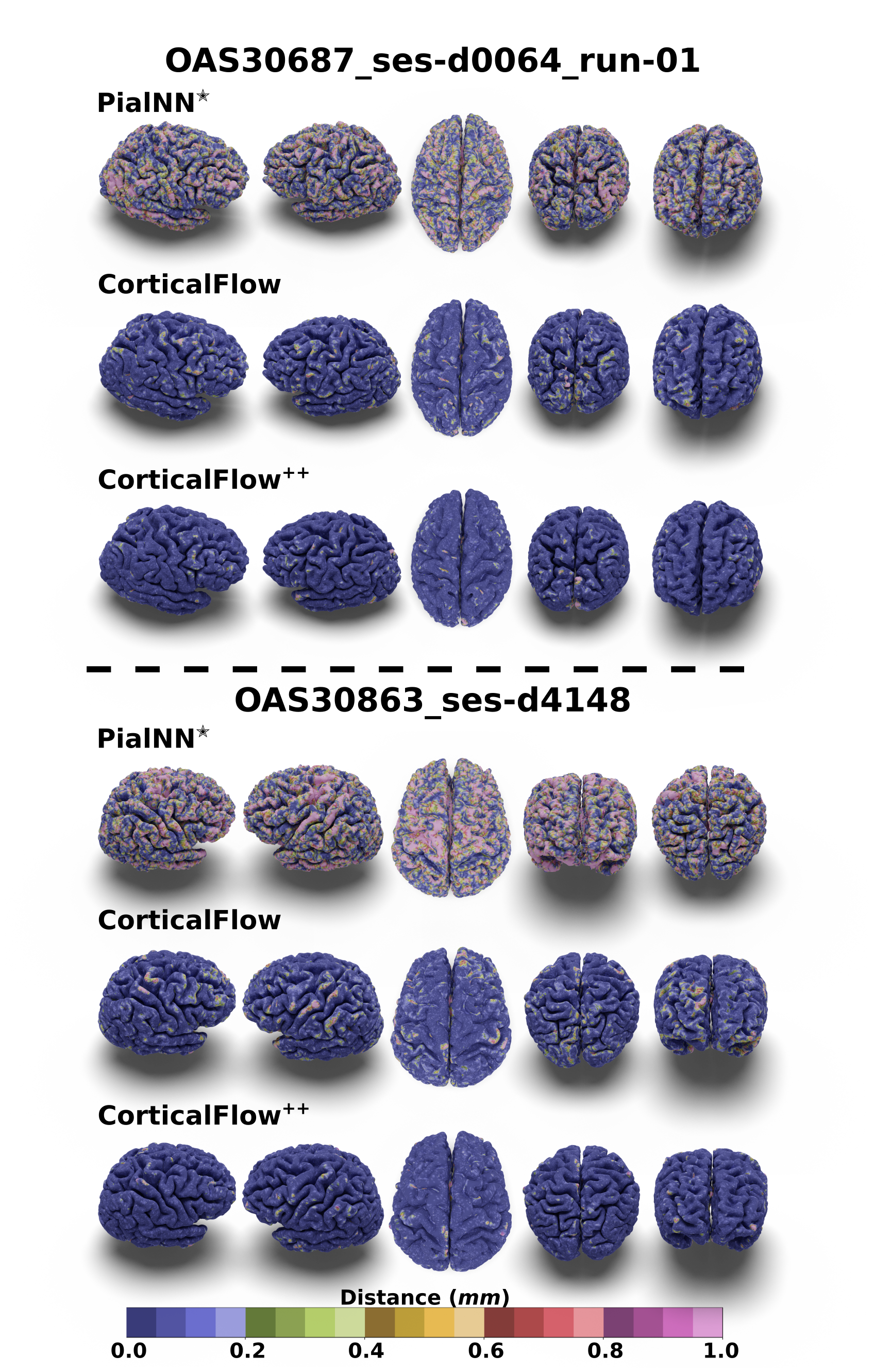} 
    \caption{Predicted cortical surfaces for \textbf{OASIS3} subjects. The surfaces are color-coded with the distance to the pseudo-ground-truth surfaces.}
\end{figure}

\clearpage

%
%
%
\bibliographystyle{splncs04}
\bibliography{biblio}
%




\end{document}